\documentclass{article}
\usepackage[T1]{fontenc}
\usepackage[utf8]{inputenc}
\usepackage[protrusion=true, expansion=true, tracking=small]{microtype}
\usepackage[]{ismir} 
\usepackage{amssymb}
\usepackage{amsmath,cite,url}
\usepackage{graphicx}
\usepackage{color}

\title{Refining music sample identification with a self-supervised graph neural network}

\multauthor
{Aditya Bhattacharjee$^1$ \hspace{0.4cm} Ivan Meresman Higgs$^1$ \hspace{0.4cm} Mark Sandler$^1$ \hspace{0.4cm} Emmanouil Benetos$^1$}
{\begin{minipage}{\textwidth}\centering
 $^1$ Queen Mary University of London, UK\\
 {\tt\small \{a.bhattacharjee, i.meresman-higgs, mark.sandler, emmanouil.benetos\}@qmul.ac.uk}
 \end{minipage}
}

\def\authorname{A. Bhattacharjee, I. Meresman Higgs, M. Sandler, and E. Benetos}

\begin{document}
\maketitle

\begin{abstract}
Automatic sample identification (ASID) - the detection and identification of portions of audio recordings that have been reused in new musical works - is an essential but challenging task in the field of audio query-based retrieval. While a related task, audio fingerprinting, has made significant progress in accurately retrieving musical content under ``real world'' (noisy, reverberant) conditions, ASID systems struggle to identify samples that have undergone musical modifications. Thus, a system robust to common music production transformations such as time-stretching, pitch-shifting, effects processing, and underlying or overlaying music is an important open challenge. 
In this work, we propose a lightweight and scalable encoding architecture employing a Graph Neural Network within a contrastive learning framework. Our model uses only 9\% of the trainable parameters compared to the current state-of-the-art system while achieving comparable performance, reaching a mean average precision (mAP) of 44.2\%.

To enhance retrieval quality, we introduce a two-stage approach consisting of an initial coarse similarity search for candidate selection, followed by a cross-attention classifier that rejects irrelevant matches and refines the ranking of retrieved candidates - an essential capability absent in prior models. In addition, as 
queries in real-world applications are often short in duration, we benchmark our system for short queries using new fine-grained annotations for the Sample100 dataset, which we publish as part of this work.
\end{abstract}

\section{Introduction}

\textit{Sampling} is a musical technique that ``incorporates portions of existing sound recordings into a newly collaged composition'' \cite{mcleod_2011}. The samples often undergo significant modification during this creative process: they may be pitch-shifted, time-stretched and heavily processed with audio effects (henceforth \textit{sampling transformations}), and are typically combined with other musical elements, creating ``musical interference'' which makes identification difficult even for human experts. 
The relevance of this practice is highlighted as, since the mass popularisation of hip hop, disco and electronic dance music, this kind of ``transformative appropriation'' has become one of the most important techniques for composers and songwriters\cite{demers_2006}.

Automatic sample identification (ASID) is a crucial task in music retrieval: given an audio query - either a small segment or an entire music track - the goal is to retrieve the sample source from a database of music recordings, even if sampling transformations have been applied. The potential to substantially impact domains such as attribution and copyright highlights the relevance of this task for music creators and rights holders, as well as music information retrieval (MIR) researchers.

This task is particularly challenging as sampling transformations can drastically alter the audio features while maintaining perceptual similarity. A reasonable approach is to take cues from deep learning-based audio fingerprinting research, learning metrics that allow for a similarity-based search and retrieval system. Additionally, augmentations in the training pipeline allow models to learn invariance to sampling transformations employed in music production. Recent audio fingerprinting research has successfully employed Graph Neural Networks (GNNs), achieving state-of-the-art results while using compact architectures that facilitate efficient training, which informs this work.

Progress in ASID has been hindered by the limited availability of well-annotated datasets that reflect real-world sampling practices. 
The Sample100 dataset \cite{vanbalen_2011} is the only publicly available dataset of annotations specifically addressing the presence of samples in commercially produced songs. In this paper we present a revised version of this dataset, annotated by experts to include more fine-grained temporal annotations of the samples, as well as additional comments, time-stretching estimates and instrumentation information. We use these new annotations to report segment-wise hit-rates and to analyse the performance of our system in relation to the type of sample and augmentations 
performed during the artistic process.


Our key contributions are as follows:
\begin{itemize}
\setlength\itemsep{0.1em}
\item We propose the adaptation of a lightweight Graph Neural Network as the neural encoder for ASID.
\item We introduce a binary cross-attention classifier to facilitate an accurate ranking and refining of retrieved audio fingerprints.
\item We contribute new fine-grained temporal annotations to the Sample100 dataset, and evaluate our model’s performance on short-query retrieval, demonstrating superior top-N hit-rates compared to the baseline.
\item We present a detailed analysis of retrieval performances on different types of samples and discuss the viability of the proposed framework.     
\end{itemize}
Our code as well as the newly extended Sample100 dataset have been made available for reproducibility\footnote{https://github.com/chymaera96/NeuralSampleID}.


\section{Related Works}
Despite the ASID task being a relevant and challenging one for the MIR community, there have been few attempts to tackle it. Foundational work by Van Balen et al. \cite{vanbalen_2011}, introduced the Sample100 dataset and proposed the adaptation of a spectral peak-based audio fingerprinting framework to make it robust to pitch-shifting. Gururani et al. \cite{gururani_2017} proposed a system inspired by music cover identification, using Non-negative Matrix Factorization to create templates of the samples and Dynamic Time Warping to achieve a detection algorithm that could be robust to time-shifting. Both of these works focus primarily on robustness against individual sampling transformations 
but neither address the broader range 
typically encountered in real-world 
scenarios. Other traditional fingerprinting methods that were effective for audio retrieval tasks, such as audfprint \cite{ellis_2014} and Panako \cite{six_2014} have also been tested on this task \cite{cheston_2025} and proved insufficient for ASID, struggling with the challenges of combined sampling transformations and interfering ``musical noise'' (the overlying musical composition).


More recently, the first deep learning-based approach by Cheston et al. \cite{cheston_2025} achieved state-of-the-art performance on the Sample100 dataset using a CNN architecture (ResNet50-IBN) previously used for cover song identification \cite{du_2021} and exploiting music source separation to create synthetic training data. This approach serves as our baseline and demonstrates both the feasibility and remaining challenges of applying deep learning to ASID.

Current state-of-the-art audio retrieval systems predominantly use CNNs \cite{du_2021, xu_2018, chang_2021} or transformers \cite{singh_2022} trained with contrastive learning objectives. While effective, these architectures typically require significant computational resources and large training batches, limiting their practical viability. These limitations can be addressed by more parameter-efficient approaches based on Graph Neural Networks (GNNs), which excel at capturing complex structural patterns in non-Euclidean spaces \cite{li_2019}. GNNs have proven effective for audio tasks where temporal and spectral relationships are important, including audio fingerprinting \cite{grafp} and audio tagging \cite{singh_2024}, by effectively modelling local and global interactions between time-frequency regions.

\section{Methodology}
\label{section:methodology}

\begin{figure*}[]
    \centering
    \includegraphics[trim=0.5cm 22.8cm 1.2cm 1cm, clip, width=\linewidth]{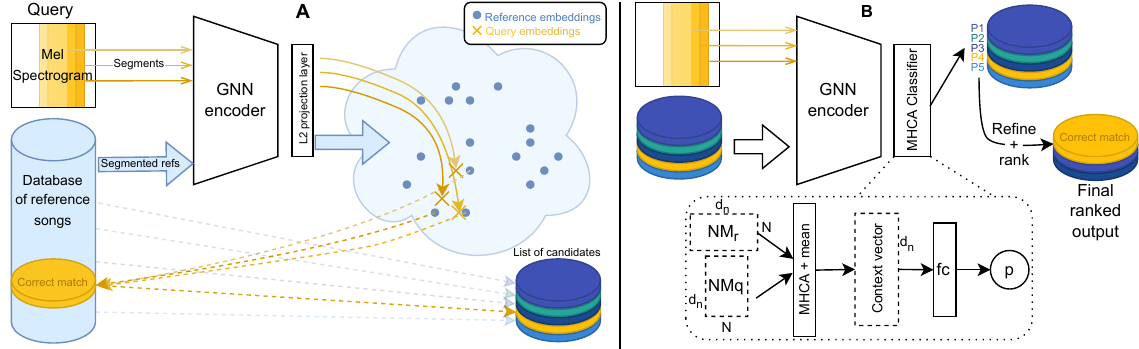}
    \caption{Illustrated ASID methodology: (A) Given a query, we compute segment-level embeddings (fingerprints), matched to reference embeddings via approximate nearest-neighbour (ANN) search; based on which, candidate songs are retrieved from the reference database through a lookup process (dotted arrows). (B) A multi-head cross-attention (MHCA) classifier refines and ranks candidates using node embedding matrices $\text{NM}_q$ (query) and $\text{NM}_r$ (references).}
    \label{fig:eval-pipeline}
\end{figure*}

ASID involves two categories of audio recordings: a \textit{reference}, an original music recording, and a \textit{query}, a new recording that incorporates (i.e., samples) parts of the reference. For training, we generate query-reference pairs by re-mixing source separated stems as proposed in \cite{cheston_2025}. For evaluation, our retrieval methodology employs a two-stage process: initial candidate selection via approximate nearest-neighbour search, followed by fine-grained ranking with the cross-attention classifier. Figure~\ref{fig:eval-pipeline} illustrates the complete retrieval pipeline, detailing how reference matches are retrieved and ranked for a given query.

\subsection{Input Features}

Our system employs log-scaled Mel-spectrograms as input features. Given an audio waveform $ y \in \mathbb{R}^t $, sampled at 16 kHz, we first compute its Mel-spectrogram representation $ \mathcal{X} \in \mathbb{R}^{F \times T} $. Here, $ F $ denotes the number of Mel-frequency bins, and $ T $ is the number of temporal frames.

During training, we randomly sample short audio segments of fixed duration $ t_{\text{seg}} $ from each recording in the training dataset and use it to generate proxy query-reference pairs (see Section 3.4.1). For retrieval, we use real query and reference audio recordings which are segmented into overlapping segments of length $ t_{\text{seg}} $. Section \ref{sec:config} details the configuration of the input features and hyperparameters.

\subsection{Encoder Architecture}

Our GNN encoder builds upon the architecture introduced in~\cite{grafp}. Given an input spectrogram \(\mathcal{X}\), we first represent it as a set of three-dimensional time-frequency points, each described by its time index, frequency bin index, and amplitude value. From this initial representation, we produce overlapping \textit{patch embeddings} by aggregating local neighbourhoods of time-frequency points into latent vectors. Formally, each resulting patch embedding is represented by:
\begin{equation}
f : \mathbb{R}^{3 \times p} \rightarrow \mathbb{R}^{d},
\end{equation}
where \(p\) denotes the number of neighbouring points aggregated per patch, and \(d\) is the dimensionality of the latent embedding. These patch embeddings serve directly as nodes in the subsequent graph structure.

Next, we construct a k-nearest neighbour (kNN) graph from these node embeddings. Specifically, for each node embedding \( x_i \), we identify its \( k \) nearest neighbours based on cosine similarity in the latent embedding space. The resulting edges represent latent structural relationships among spectrogram patches.

Node embeddings are then iteratively refined via graph convolution (\textit{GraphConv}) layers. For each node embedding \(x_i\), we aggregate information from its neighbours \(x_j\), where \(j \in \mathcal{N}(x_i)\). Formally, the update rule is given by:
\begin{equation}
y_i = x_i + \sigma\bigl(\text{AGG}(\{x_j : j \in \mathcal{N}(x_i)\})\bigr),
\end{equation}
where \(y_i\) is the updated embedding, \(\sigma\) denotes a nonlinear activation function, \(\mathcal{N}(x_i)\) is the set of neighbours of node \(x_i\), and AGG represents an aggregation operation summarizing relevant information from neighbouring nodes. Through iterative aggregation, each node embedding progressively encodes increasingly rich contextual and structural information. The GNN encoder comprises multiple blocks of \textit{GraphConv} layers, each followed by feedforward network (\textit{FFN}) layers. At the beginning of each block, the kNN graph is dynamically reconstructed to reflect the updated node embeddings. 

The output of the GNN encoder is a set of refined node embeddings, collectively referred to as the \textit{node embedding matrix}, which serve as input features to the cross-attention classifier. Finally, these node embeddings are average-pooled and projected into \textit{audio fingerprints}. Both latent embeddings are  used in the subsequent retrieval refinement stage.

For a comprehensive discussion of architectural details and design considerations, we refer readers to~\cite{grafp}.

\subsection{Cross-Attention Classifier}


 To capture the latent relationships between these two sets of node embeddings, we introduce a multi-head cross-attention classifier. Given a query and a reference audio segment, we first compute the node embedding matrices $q \in \mathbb{R}^{N \times d_n}$  and $ r \in \mathbb{R}^{N \times d_n} $, respectively. Here, $ N $ is the number of nodes, and $ d_n $ is the dimensionality of each node embedding. We compute attention-weighted embeddings as follows:
\begin{equation}
    \mathbf{C} = \text{MHA}(q, r, r)
\end{equation}
where $\text{MHA}(.)$ denotes standard multi-head attention~\cite{vaswani2017attention}. The resulting embedding matrix $\mathbf{C} \in \mathbb{R}^{N \times d_n}$ is an attention-weighted transformation of $r$, where attention is computed between corresponding node embeddings in $q$ and $r$. $\mathbf{C}$ is then aggregated by mean pooling, producing a single context vector \(\mathbf{c} \in \mathbb{R}^{d_n}\):
\begin{equation}
    \mathbf{c} = \frac{1}{N}\sum_{j=1}^{N}\mathbf{C}_j
\end{equation}

where $C_j$ is context vector of the $j$-th node embedding. Finally, the context vector $\mathbf{c}$ is transformed by a shallow nonlinear classifier into a scalar confidence score $s$:
\begin{equation}
    s = \sigma(\mathbf{w}^T\mathbf{c} + b)
\end{equation}
where $\mathbf{w} \in \mathbb{R}^{d_n} $, $ b \in \mathbb{R}$ are learnable parameters, and $\sigma$ denotes the sigmoid activation function. The scalar $s$ indicates the confidence that the query and reference segments match. As shown in Figure \ref{fig:eval-pipeline}, at retrieval time, this is used as a ranking mechanism as well as a measure for rejecting low-confidence candidates.

\subsection{Training Pipeline}

Our proposed approach involves two distinct training stages: a self-supervised contrastive learning stage for embedding training and a subsequent binary classification stage for the downstream cross-attention classifier. Both stages use identical procedures to produce proxy query-reference pairs from the source-separated training data, closely following the methodology established in prior work~\cite{cheston_2025}.

\subsubsection{Query-Reference Pair Generation}

Let us denote the stems extracted from the training audio source \( x \) as a set \(\mathcal{S} = \{s_1, s_2, ..., s_K\}\), where each stem \(s_k\) corresponds to a source-separated audio component (e.g., vocals, drums, bass). Given a random timestamp segment \( t_s \) starting at \( t \) and of length \(\Delta t\), we first extract corresponding audio segments from each stem as
\begin{equation}
    s_k(t_s) = s_k[t, t + \Delta t]
\end{equation}
resulting in the set \(\{s_1(t_s), s_2(t_s), ..., s_K(t_s)\}\). These stem segments are partitioned randomly into two subsets, \(\mathcal{S}_{q}\) and \(\mathcal{S}_{r}\), with 
\(\mathcal{S}_{q} \cup \mathcal{S}_{r} = \mathcal{S} \) and \(\quad \mathcal{S}_{q} \cap \mathcal{S}_{r} =\varnothing\).

A query segment \( x_q \) is formed as the sum of stems in \(\mathcal{S}_{q}\):
\begin{equation}
x_q = \sum_{s \in \mathcal{S}_{q}} s(t_s).
\end{equation}

The reference segment \( x_r \) is generated by mixing an augmented version of the query segment with the remaining stems:
\begin{equation}
x_r = \text{aug}_2\left(\text{aug}_1\left(x_q\right) + \sum_{s \in \mathcal{S}_{r}} s(t_s)\right).
\end{equation}

Here, \(\text{aug}_1\) and \(\text{aug}_2\) represent audio effects functions applied sequentially to simulate realistic music production transformations. The effect parameters are sampled from a uniform distribution. Specifically, 
\begin{itemize}
    \item \textit{\(\text{aug}_1\)}: time-offset ($\pm$250ms) and gain variation ($\pm$10dB).
    \item \textit{\(\text{aug}_2\)}: pitch-shifting ($\pm$3 semitones) and time-stretching (70 - 150\%).
\end{itemize}

The source-separation system (see Section 4.1) allows the extraction of musically salient sources that can constitute a sample. The pair \( (x_q, x_r) \) constitutes a positive query-reference example; \(x_q\) is a proxy for a query containing an instance of a sample, and \(x_r\) represents a reference example which contains the sample that is creatively distorted and is present in a mix along with other musical elements.

\subsubsection{Contrastive Learning}

We train the encoder using a self-supervised contrastive learning framework. Given a batch of \(B \) pairs \(\{(x_q^i, x_r^i)\}_{i=1}^{B}\), we obtain their corresponding audio fingerprints \(\{z_q^i, z_r^i\}_{i=1}^{N}\) from the encoder. We then employ the Normalized Temperature-scaled Cross Entropy (NT-Xent) loss~\cite{chen2020simple} to maximize similarity between embeddings from positive pairs, while minimizing cosine similarity to embeddings from all other pairs in the batch.

\subsubsection{Downstream Classifier Training}

The cross-attention classifier is trained as a downstream task, with the encoder parameters frozen after the contrastive learning stage. For this stage, we discard the previously used projection network and directly use the node embedding matrix obtained from the encoder.

Training batches consist of query-reference pairs generated identically to the contrastive learning stage. Let \(\mathcal{Q} = \{q_i\}_{i=1}^{B_c}\) and \(\mathcal{R} = \{r_j\}_{j=1}^{B_c}\) represent query and reference embedding sets in a batch, respectively, where each embedding \(q_i, r_j \in \mathbb{R}^{N \times d_n}\), and \(B_c\) is the batch size. 

Positive examples correspond to pairs of identical indices:
\begin{equation}
\mathcal{P} = \{(q_i, r_j)\mid i = j\},
\end{equation}
while negative examples are selected from pairs with non-identical indices via hard-negative mining. Specifically, we select negative pairs as the subset of non-positive pairs that maximize audio fingerprint similarity, thus being the most confounding:
\begin{equation}
\mathcal{N} = \{(q_i, r_j^-)\mid i \neq j,\, r_j^- = \arg\max_{r_j, j\neq i}\text{sim}(z_i,z_j)\}.
\end{equation}

We maintain a fixed ratio of 1:3 for positive to negative pairs within each training batch. The classifier outputs a scalar prediction \(p \in [0,1]\), trained with the binary cross-entropy (BCE) loss,
where the label for pairs \((q_i,r_j)\in \mathcal{P}\) is 1 and for for pairs \((q_i,r_j^-)\in \mathcal{N}\) is 0. 

\subsection{Retrieval and Evaluation}

Our retrieval system, illustrated in Figure \ref{fig:eval-pipeline}, operates in two sequential stages:
\begin{itemize}
\setlength\itemsep{0.1em}
    \item \textit{Approximate nearest-neighbour (ANN) search} to do a fast and coarse search of candidate reference audio fingerprints from the database.
    \item \textit{Cross-attention classifier scoring} to refine the candidate set and rank them based on relevance. 
\end{itemize}

For every overlapping segment (computed as described in Section 3.1) in the query, we probe the reference database for matches based on the similarity of the audio fingerprints; thus yielding a set of candidate matches. 

In the second stage, we utilize the cross-attention classifier to refine these candidate matches. For each candidate segment retrieved, we extract its corresponding node embedding matrix. Given a query recording, represented as a sequence of node embedding matrices, we compute classifier scores \(p(q, r)\) for each pair of query \(q\) and retrieved candidate \(r\). The final candidate segment-level confidence score is determined by selecting the maximum classifier score over all segments of the query:
\begin{equation}
p_\text{clf}(q, r) = \max_{q_i \in Q} p(q_i, r).
\end{equation}

We reject candidate segments with confidence scores \( p_\text{clf}(q,r) < 0.5 \). Subsequently, we aggregate these accepted segment-level scores to obtain a song-level retrieval score. Specifically, for each unique reference recording, we sum the segment-level confidence scores:
\begin{equation}
P_\text{song}(q, R) = \sum_{r \in R}P_\text{clf}(q, r),
\end{equation}
where \(R\) denotes the set of retrieved segments belonging to the same reference song. The resulting aggregated scores \(P_\text{song}(q,R)\) provide a robust ranking of candidate songs for each query recording.

\section{Dataset}
\label{section:dataset}
\subsection{Training Dataset}

For training, we use the Free Music Archive (FMA) \textit{medium} dataset \cite{fma_2017}, which contains 25,000 30-second tracks across 16 genres. We pre-processed this dataset to make it suitable for our stem-mixing contrastive learning approach. 
We used the current SOTA algorithm ``BeatThis'' \cite{foscarin_2024} to perform beat tracking and use this as a proxy for musical rhythmic regularity in the FMA tracks, excluding 2,533 
tracks with fewer than 32 beats after the first downbeat. This filtering ensured that our training data consisted only of musical content with some level of rhythmic structure.

To generate the stems that will be used for the synthetic training pairs, we applied source separation using the Hybrid Transformer Demucs model (\texttt{htdemucs})~\cite{rouard_2022} to each usable track, separating them into drums, bass, vocals, and ``other'' stems.

\subsection{Evaluation Dataset}


For the evaluation of our system, we use the Sample100 dataset \cite{vanbalen_2011}
. The dataset consists of 75 full-length hip-hop recordings (queries) containing samples from 68 full-length songs (references) across a variety of genres, with R\&B/Soul representing the majority \cite{cheston_2025}. It contains 106 sample relationships and a total of 137 sample occurrences, as some queries use multiple samples and some references appear in multiple queries. To challenge retrieval systems, the dataset includes 320 additional ``noise'' tracks with a similar genre distribution, which are not sampled in any query. 


Because samples are typically created from a short segment of a song, 
only a small portion of each candidate track is sampled and present in queries - sample lengths range from just one second to 26 seconds. The samples represent real-world musical ``transformative appropriation'' \cite{demers_2006}, including both tonal (\textit{riffs}), percussive drum break (\textit{beats}), and \textit{1-note} micro-samples. 
Non-musical samples (e.g. film dialogue) are not included.

To enable more detailed evaluation, we present an extended version of the Sample100 dataset with fine-grained temporal annotations performed by expert musicians using Sonic Visualiser \cite{cannam_2010}. Unlike the original dataset, which only provided first occurrence timestamps at 1-second precision, our annotations include precise start and end times for all sample occurrences with $\pm250$ms resolution, transforming the dataset into a segment-wise evaluation resource. This improved temporal granularity allows for more accurate evaluation of ASID systems by testing with short query snippets from anywhere within the sampled material.

We further enrich the dataset by adding estimates of the time-stretching ratio between the reference and query tracks, as well as instrumentation (stem) annotations for both the original material and the interfering instruments in the query, and expanding the comments about the samples. The time-stretching ratio was calculated from the tempo of both query and reference segments, determined through a combination of automatic beat tracking \cite{foscarin_2024} with manual verification. Stem annotations were performed by listening to the tracks and their source-separated stems to ensure accuracy. 
Relevant sample class counts are shown in Table \ref{tab:sample-cat-performance}, including a 
categorisation into substantial or minimal time-stretching. 
This new information will enable more nuanced analysis of our model's performance across different types of sampling practices in section \ref{section:results-ablation}.



\section{Experimental Setup}
\label{sec:config}

\subsection{Hyperparameters and Configuration}
Our experimental setup and hyperparameter choices are summarized in Table~\ref{tab:hyper}, with certain parameters detailed in the preceding sections. The contrastive learning stage was performed using an NVIDIA A100 GPU, with models trained for a maximum of 180 epochs; we employed early stopping based on validation performance. Training utilized the Adam optimizer coupled with a cosine annealing learning-rate scheduler.  For the downstream cross-attention classifier, we trained for a maximum of 5 epochs using the Adam optimizer with a fixed learning rate, keeping the encoder parameters frozen to preserve the learned representations from the contrastive learning stage. For the ANN search algorithm, we use IVF-PQ \cite{johnson2019billion}, an efficient choice for retrieval tasks in large vector databases.

\begin{table}[h]
    \centering
    \resizebox{0.85\columnwidth}{!}{%
    \begin{tabular}{|l|l|}
        \hline
        \textbf{Hyperparameter} & \textbf{Value} \\
        \hline
        Sampling rate & 16,000 Hz \\
        log-power Mel-spectrogram size $F \times T$ & $64\times 32$ \\
        Fingerprint \{window length, hop\} & \{4s, 0.5s\} \\
        Fingerprint dimension & 128 \\
        Node matrix dimension \{$N$, $d_n$\} & \{32, 512\} \\
        Temperature $\tau$ & 0.05 \\
        Contrastive batch size $B$ & 1024 \\
        Downstream batch size $B_c$ & 32 \\
        \hline
    \end{tabular}
    }
    \caption{Experimental Configuration}
    \label{tab:hyper}
\end{table}

\subsection{Evaluation Metrics}
\label{section:evaluation}

The ASID task is fundamentally a retrieval problem, where the goal is to rank candidate audio segments based on their relevance to a query. Hence, we adopt mean average precision (mAP) \cite{downie2008music} as our primary metric, where the query is computed from a full song containing a sample. The mean average precision (mAP) summarizes retrieval quality by aggregating the precision values at the ranks where relevant items are retrieved, averaged across all queries. 

Additionally, inspired by an established practice in audio fingerprinting literature~\cite{chang_2021}, we report top-$N$ hit rates. Specifically, we measure the proportion of queries for which at least one correct sample is retrieved within the top $N$ ranked results. We do so for different query sizes (5s to 20s). This metric provides an intuitive indication of practical retrieval accuracy and the system's efficacy for short queries.

\subsection{Baseline Framework}

We compare our proposed system against the recent state-of-the-art baseline introduced by Cheston et al.~\cite{cheston_2025}. Their framework employs a ResNet50-IBN architecture and utilizes a multi-task learning approach that jointly optimises a metric learning objective through triplet loss and an auxiliary classification task. This architecture has achieved state-of-the-art retrieval performance in terms of mean average precision (mAP). Due to practical computational constraints, we instead adopt and report results on a ResNet18-IBN model, which has a comparable number of parameters to our proposed GNN-based encoder. Apart from the model size, we closely adhere to the training procedures and evaluation methodology outlined in \cite{cheston_2025}. We also include their reported best performance for reference.

\section{Results and Discussion}
\label{section:results}
\subsection{Benchmarking}

We present the performance comparison between our proposed GNN+MHCA architecture and the baseline in Table~\ref{tab:benchmarking}. Our model matches the reported performance of the much larger ResNet50-IBN model, and significantly outperforms the reimplemented baseline.

\begin{table}[h!]
\centering
\resizebox{0.9\columnwidth}{!}{%
\begin{tabular}{|ll|c|c|}
\hline
\multicolumn{2}{|c|}{\textbf{Model}} & \textbf{\# params} & \textbf{mAP} \\ \hline
\multicolumn{2}{|l|}{ResNet50-IBN (Cheston et al.)} & 222M & 0.441 \\ \hline
\multicolumn{2}{|l|}{ResNet18-IBN (Baseline)} & 34M & 0.330 \\ \hline
\multicolumn{1}{|l|}{GNN (Ours)} & batch size = 1024 & 20M & 0.416 \\ \hline
\multicolumn{1}{|l|}{GNN +} & batch size = 256 &  & 0.373 \\ \cline{2-2} \cline{4-4} 
\multicolumn{1}{|l|}{MHCA} & batch size = 512 & 20M & 0.411 \\ \cline{2-2} \cline{4-4} 
\multicolumn{1}{|l|}{(Ours)} & batch size = 1024 &  & \textbf{0.442} \\ \hline
\end{tabular}%
}
\caption{Performance of models on Sample100 dataset.}
\label{tab:benchmarking}
\end{table}

A key factor influencing our model's performance is the batch size used during contrastive learning. Increasing the batch size from 256 to 1024 leads to an improvement of 6.9pp in mean average precision (mAP). This improvement occurs because larger batch sizes provide more negative samples per positive pair, enriching the diversity of the contrastive space. Consequently, the model learns embeddings that better discriminate between relevant and irrelevant examples.

\begin{table}[h!]
\centering
\resizebox{\columnwidth}{!}{%
\begin{tabular}{|cccccc|}
\hline
\multicolumn{1}{|c|}{\textbf{Models}} & \multicolumn{5}{c|}{\textbf{Query Length}} \\ \hline
\multicolumn{1}{|c|}{} & 5 sec & 7 sec & 10 s & 15 sec & 20 sec \\ \hline
\multicolumn{6}{|c|}{Top-1 Hit Rate (\%)} \\ \hline
\multicolumn{1}{|c|}{ResNet18-IBN (Baseline)} & \textbf{15.8} & 15.8 & 16.0 & 16.5 & 21.7 \\ \hline
\multicolumn{1}{|c|}{GNN+MHCA (Ours)} & 15.5 & \textbf{24.6} & \textbf{27.5} & \textbf{26.1} & \textbf{30.7} \\ \hline
\multicolumn{6}{|c|}{Top-3 Hit Rate (\%)} \\ \hline
\multicolumn{1}{|c|}{ResNet18-IBN (Baseline)} & \textbf{24.3} & 24.3 & 29.2 & 26.1 & 32.3 \\ \hline
\multicolumn{1}{|c|}{GNN+MHCA (Ours)} & 19.1 & \textbf{32.1} & \textbf{38.3} & \textbf{40.9} & \textbf{50.3} \\ \hline
\multicolumn{6}{|c|}{Top-10 Hit Rate (\%)} \\ \hline
\multicolumn{1}{|c|}{ResNet18-IBN (Baseline)} & \textbf{27.4} & 27.4 & 40.4 & 45.2 & 49.1 \\ \hline
\multicolumn{1}{|c|}{GNN+MHCA (Ours)} & 19.1 & \textbf{33.8} & \textbf{44.7} & \textbf{51.3} & \textbf{63.2} \\ \hline
\end{tabular}%
}
\caption{Hit rates of our framework and baseline.}
\label{tab:hit-rates}
\end{table}

Table \ref{tab:hit-rates} shows our model's performance on short queries, a common use case in real-world sample identification scenarios. While the hit rates for shorter queries are comparable to the baseline, our framework exhibits significantly superior performance for longer queries (14.1pp increase in top-1 hit rate for 20-second-long queries). The progressive improvement in hit rates with increasing query length show that our approach effectively aggregates segment-level confidence scores to retrieve the correct reference song.

\subsection{Retrieval Refinement via Cross-Attention Classifier}
To examine the impact of the cross-attention classifier as a retrieval refinement step, we conduct an ablation study. Table~\ref{tab:benchmarking} shows that incorporating the classifier (MHCA) to rank retrieved results improves mAP by 2.6pp, confirming the utility of this refinement stage. Additionally, to evaluate the classifier's capability to reject irrelevant matches, we construct a balanced validation set comprising 300 positive query-reference pairs and 300 negative pairs drawn from the ``noise'' data described in Section~4.2. The classifier achieves an AUROC score of 0.776, indicating that it does not perfectly discriminate between genuine and confounding examples. Thus, the observed improvement in retrieval performance can be attributed to the combined effect of the two-stage retrieval process, rather than solely to the rejection capability of the classifier.

\subsection{Performance by Sample Characteristics}
\label{section:results-ablation}
To understand the performance of the model across sample characteristics, we computed the mAP for different categories of Sample100. As shown in Table \ref{tab:sample-cat-performance}, there is a modest performance gap between melodic/harmonic \textit{riff} samples and percussive \textit{beat} samples (the two \textit{1-note} samples were not taken into account). This may be attributed to the nature of beat samples, which consist primarily of drums that are often subject to overdubbing techniques where producers layer additional percussion elements, and that might also be buried deeper in the mix beneath other instrumentation, potentially making them less salient for the GNN to capture. Further analysis by looking at the specific instrumentation in the reference and query, or by applying source separation at detection time, is left for future work.

\begin{table}[h!]
\centering
\resizebox{0.85\columnwidth}{!}{%
\begin{tabular}{|c|ccc|cc|}
\hline
 & \multicolumn{3}{c|}{\textbf{Type}} & \multicolumn{2}{c|}{\textbf{Time stretching}} \\ \hline
\textbf{} & \multicolumn{1}{c|}{Riff} & \multicolumn{1}{c|}{Beat} & 1-note & \multicolumn{1}{c|}{\textgreater{}5\%} & \textless{}5\% \\ \hline
\textbf{\# samples} & \multicolumn{1}{c|}{71} & \multicolumn{1}{c|}{33} & 2 & \multicolumn{1}{c|}{44} & 62 \\ \hline
\textbf{mAP} & \multicolumn{1}{l|}{0.471} & \multicolumn{1}{l|}{0.391} & - & \multicolumn{1}{c|}{0.340} & 0.503 \\ \hline
\end{tabular}%
}
\caption{Performance according to sample class.}
\label{tab:sample-cat-performance}
\end{table}


A significant performance gap was observed in relation to time stretching, where we classified samples subjected to minimal time stretching (<5\%) and those with significant time stretching (>5\%). This 16.3pp difference in performance shows that although our model is robust to some degree of time-stretching, large changes in tempo fundamentally alter the temporal relationships between audio features that our model relies on for identification.



\section{Conclusion and Future Work}
This paper presents a lightweight GNN-based approach for automatic sample identification that achieves state-of-the-art performance while using only 9\% of the parameters compared to previous methods. Our key contributions include adapting a GNN encoder for sample identification, introducing a cross-attention classifier for refining retrieval results, and extending the Sample100 dataset with fine-grained temporal annotations that enable granular evaluation.

Our results show that the proposed framework achieves a mAP of 44.2\%, with strong performance on melodic-harmonic samples and samples with low time-stretching. Our framework's cross-attention stage is useful in the refining of the ranking, and introduces rejection capabilities.

Future work should explore end-to-end training methods robust to sampling transformations, and explore integrating source separation during inference to improve performance on heavily masked samples. Newly available annotations can be leveraged for analysis of how specific attributes of samples (instrumentation type, interpolation, genre) affect identification accuracy, and we hope the release of our extended Sample100 dataset will aid the development of specialized techniques that address the most challenging cases in ASID.



\section{Acknowledgments}
We thank Matthew Alan Walford for his contributions to the new annotations of the Sample100 dataset.
This research utilised Queen Mary's Apocrita HPC facility, supported by QMUL Research-IT (http://doi.org/10.5281/zenodo.438045).
This work was supported by UKRI - Innovate UK (Project no. 10102241).
A. Bhattacharjee is a research student at the UKRI Centre for Doctoral Training in Artificial Intelligence and Music, supported jointly by UK Research and Innovation [grant number EP/S022694/1] and Queen Mary University of London.\\

\bibliography{ISMIRtemplate}

%
%
%
%

\end{document}